# The Structural Difference Between Strong and Fragile Liquids

Gang Sun[1,2], Linwei Li[1] and Peter Harrowell[1,*]

*1, School of Chemistry, University of Sydney, Sydney New South Wales 2006, Australia*

*2, Department of Fundamental Engineering, Institute of Industrial Science, University of Tokyo, 4-6-1 Komaba, Meguro-ku, Tokyo 153-8505, Japan.*

* Corresponding author: peter.harrowell@sydney.edu.au

Abstract

A structural order parameter for disordered configurations is defined, based, not on local topologies, but on the degree of local restraint imposed on each atom. This restraint parameter provides a clear distinction between a strong liquid ($SiO_2$) and a fragile liquid (a binary Lennard-Jones mixture) without reference to dynamics. Where the fragile liquid exhibits an abrupt transition from unrestrained to restrain below the melting point, molten silica is highly constrained even at equilibrium. We determine the temperature dependence of the average number of particles restrained per pinned particle and consider the feature of the potential energy landscape responsible for determining the fragility.

## 1. Introduction

Having tried and discarded the heat of vaporization and the melting temperature, Laughlin and Uhlmann [1], in 1972, settled on the glass transition temperature, $T_g$, as the most satisfactory corresponding states parameter for liquid viscosities, citing, as they did so, the earlier instances of this choice of temperature scaling. The resulting plot of the log of the shear viscosity against $T_g/T$ provided a ready means to both represent and differentiate glass forming liquids with a wide range of characteristic energies in a single plot. In 1985, Austen Angell [2] took full advantage of this scaled Arrhenius plot to present an extensive compilation of glass forming liquids. In discussing the Arrhenius temperature dependence of $SiO_2$, Angell argued that this behaviour was a reflection of the underlying structure. “This structure, where every silicon is connected to four nearest neighbour silicons through bridging bonds, is self-reinforcing.” [2] Such liquids were described as ‘strong’. The deviation from Arrhenius behaviour with the addition of alkali oxides was accounted for by the relaxation of the self-reinforcement of the structure due to the disruption of the bonding network and, hence, referred to as an increase in liquid ‘fragility’. With the introduction of the strong-fragile classification of glass-forming liquids, Angell provided an important organizing concept in the study of supercooled liquids [3-5] and elevated the associated plot, now often referred to as an Angell plot, to something of an iconic status.

The point of this brief history is to establish that liquid fragility, while operationally defined by the liquid dynamics, was, as conceived by Angell, intended to describe a feature of the underlying liquid structure. If we could devise the appropriate structural order parameter directly, we could identify and characterise strong and fragile liquids by their structure alone, with no need to rely on viscosities and their variation with temperature. The realisation of this program is the goal of this paper. We present an explicit order parameter that quantifies the capacity of a static liquid configuration to restrain the motion of the constituent atoms and show that strong and fragile liquids can, indeed, be distinguished and characterised based on a structural analysis only. Given our focus on structure alone, what we are presenting in this paper is a demonstration of a clear difference in *structural* fragility between silica and an atomic alloy, as distinct from the dynamic fragility defined in terms of the temperature dependence of a relaxation time. The connection between the two types of fragility is an interesting question, part of the larger challenge of causal connection between structure and dynamics, remains a challenging problem. While we [6] and others [7] have established clear correlations between spatial heterogeneity of vibrational modes of amorphous groundstates and that of intermediate time dynamics, the problem of the microscopic account of the evolution of an amorphous configuration persists. By restricting our focus to the structural analogues of the glass forming phenomenology, we explore what progress we can make in describing at an atomic level the passage from liquid to the glassy solid while avoiding the complexities of the associated kinetics.

In this paper we shall compare the structures of a molten silica with that of a fragile liquid, a binary alloy of Lennard-Jones (LJ) particles. The local topologies of the LJ mixture, a model introduced by Kob and Andersen (KA) [8], has been studied extensively [9,10]. The topological description of a disordered close packed structure runs up against two generic difficulties. The first is that, dependent as the analysis is on identification of nearest neighbours, the majority of studies have been carried at T = 0 which is of limited use in establishing the temperature dependence of structure. The second difficulty is the large number of distinct local environments that contribute to the amorphous solid state, even at T = 0. In a recent study [11], it was found that ~ 130 different local topologies contributed significantly to the KALJ groundstate structures. Amorphous solidity, in other words, cannot generally be mapped to the presence of a just small number of local structures.

In the case of silica, the local structure is considerably better defined. The continuous random network (CRN) model, introduced by Zachariasen [12] in 1932, works quite well for T=0 glass based on the assumption that every silicon is 4-fold coordinated with oxygens and ever oxygen is bonded to two silicons. Simulation studies [13,14] have provided a more nuanced account of the defects in this ideal structure at non-zero temperatures. To understand fragility, however, we are specifically interested in significant deviations away from this idealised structure and so require a measure of structure that is not dependent on specific local geometries. As described in the following Section, we shall make use of a recently developed order parameter [15] that measures the capacity of an instantaneous configuration to restrain the motion of the constituent atoms.

The paper is organised as follows. In Section 2 we introduce the restraint order parameter and demonstrate its capacity to differentiate a strong and a fragile liquid. The structural heterogeneities of the two liquids are analysed in Section 3. In Section 4 we consider the features of the potential energy landscape responsible for the observed difference in structure.

**2. Atomic Restraint as a Structural Order Parameter**

The essential problem regarding the structure of glass forming liquids is to devise an order parameter that can be applied to each atom that measures a property of the configuration directly related to its incipient solidity. This is a challenge sidestepped in the description of thermodynamic order-disorder transitions where the appropriate order parameter can be readily crafted out the structural difference between the two phases without having to explicitly consider solidity. Recently [15], we have introduced a measure of the capacity of a configuration to restrain the motion of each constituent atom. This measure, the restraint $\mu_i$ of the *i*th atom, is defined as

$$\mu_i = \exp\left( -\frac{q^2 < \Delta r_i^2 >_{eq}}{6} \right) \quad (1)$$

where q is the magnitude of the wavevector associated with the first peak in the structure factor and $< \Delta r_i^2 >_{eq}$ is the equilibrium mean squared displacement of atom *i* using a harmonic Hamiltonian based on the normal modes of the instantaneous configuration. This calculation requires the equilibrium mean squared average of the amplitudes $A_\alpha$ of the αth normal mode. For a mode with a positive eigenvalue $\lambda_\alpha$ we can write

$$< A_\alpha^2 >= \frac{k_B T}{\lambda_\alpha} \qquad (2)$$

In the case where $\lambda_\alpha < 0$ we have calculated the analogous $< A_\alpha^2 >$ by direct numerical integration over the amplitude $A_\alpha$ (setting all other mode amplitudes to zero) using the true potential energy. For more details, the reader is referred to ref. [15]. As demonstrated previously [15], the harmonic equilibrium $< \Delta r_i^2 >_{eq}$ is less than $< \Delta r_i^2(\tau) >$, the mean squared displacement measured over some short time τ corresponding to the plateau of the self-intermediate scattering function, only approaching the kinetic values at temperatures below $0.1T_m$.

In this paper we have modelled the KA mixture using interaction parameters in ref.[8] at a composition of $A_{80}B_{20}$. The calculations were performed at a fixed number density of 1.2 with 5000 particles using constant NVT swap Monte Carlo[16]. The silica is modelled using the SHIK-1 potential described in ref. [17]. Molecular dynamics simulations were carried out for 576 particles under constant NPT with the pressure set to zero. When calculating the Hessian matrix of the second derivatives of the potential energy in silica, we used numerical differentiation due to the complications of the Ewald sum used to calculate the Coulomb energy.

In Fig. 1 we plot the average $< \mu >= \frac{1}{N}\sum_i \mu_i$ against temperature for the KA mixture and for the Si and O atoms in silica. In Fig. 2 we plot the associated distributions. The difference in the average local restraint in the two liquids is striking. In the fragile KA mixture, the restraint rises abruptly from zero at high temperatures to 1.0 (i.e. complete restraint) around a transformation temperature of T ~ 0.23. This temperature corresponds, as shown in Fig. 2, to a broad bimodal distribution between atoms with high and low restraint with the peak at higher μ first appearing around 0.6, a value that we shall associate with high restraint. Liquid silica, in contrast, exhibits this high degree of restraint across its entire temperature range, including temperatures above the melting point, i.e. the equilibrium liquid. We see no significant difference in the values of μ for the Si and O atoms, despite the difference in coordination. Structurally, silica explores the same structural range experienced by the fragile mixture over a narrower range of very low temperatures, $0 \le T / T_m \le 0.15$, over its *entire* temperature range. A strong liquid, we conclude, is one that only samples this restricted range of structures and, hence, cannot undergo the unrestrained-to-restrained transformation exhibited by the fragile liquid sub-$T_m$.

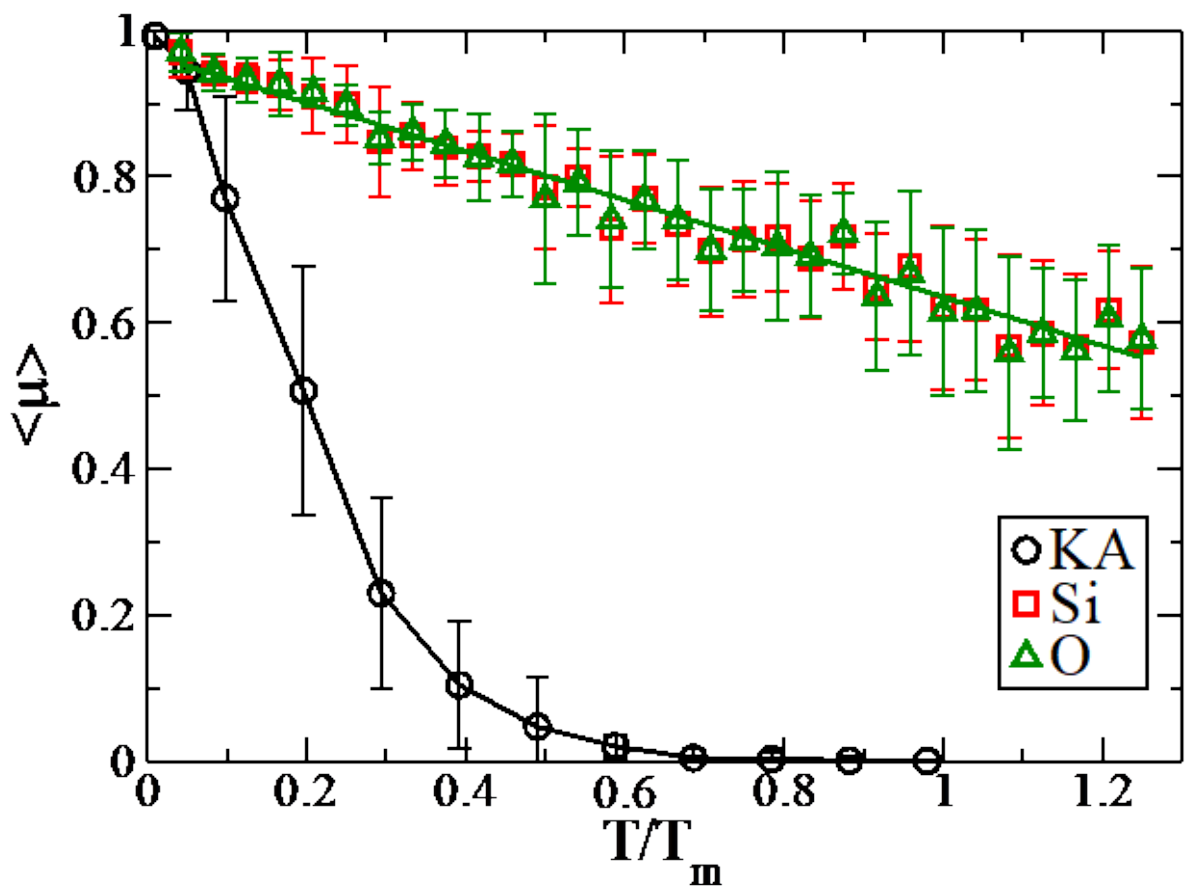

**Figure 1.** The plot of the average restraint for Si and O in liquid silica and for the KA mixture, as indicated vs $T/T_m$. The respective melting temperatures are $T_m = 1.02$ for the KA mixture [18] and $T_m = 2400K$ for silica [19].

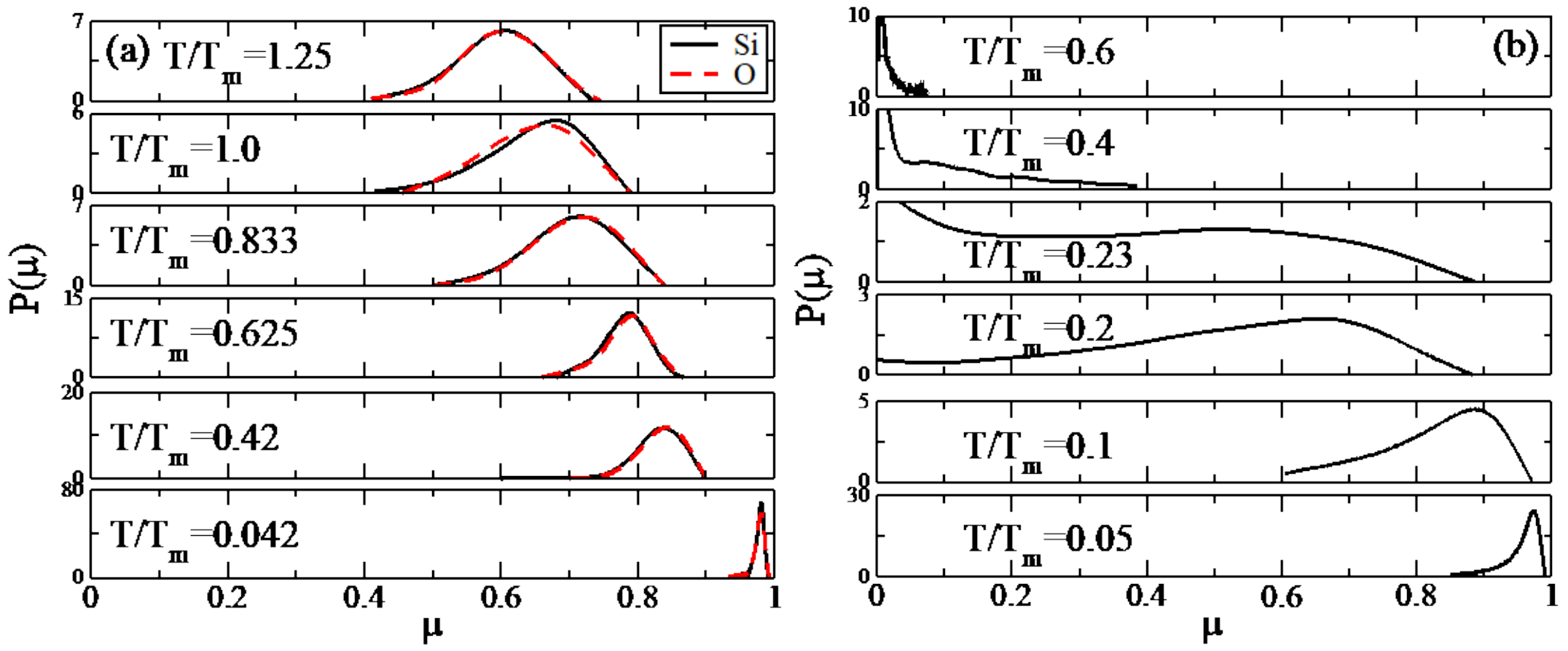

**Figure 2.** The distributions of the local restraint μ for a range of temperatures as indicated for a) Si and O in liquid silica, and b) the KA mixture.

## 3. Structural Heterogeneity and Correlation Lengths of Strong and Fragile Liquids

The structural heterogeneity of the two liquids is reflected in the width of the respective distribution of the restraint. In Fig. 3 we plot the temperature dependence of the standard deviation of μ as a function of T for silica and the binary Lennard-Jones mixture. Consistent with the distributions plotted in Fig. 2, we find that the fragile atomic mixture exhibits a peak in the heterogeneity coinciding with the transformation temperature identified above. Silica, in contrast, is considerably more uniform with a smaller standard deviation of the restraint that decreases monotonically on cooling. The difference follows directly from the presence of a structural transformation in the fragile liquid as compared with the persistently high restraint in the strong liquid across its whole temperature range.

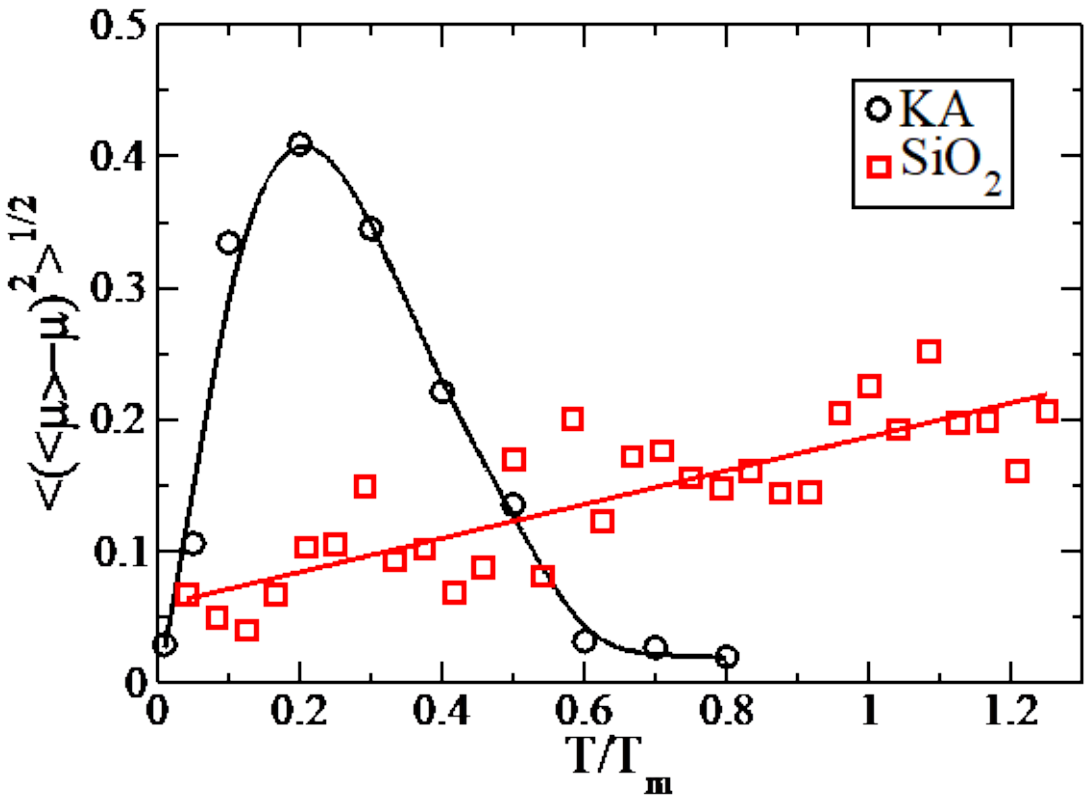

**Figure 3.** The standard deviations of the distributions of μ as a function of $T/T_m$ for silica and the KA mixture, as indicated.

Atomic restraint imparts a stiffness to the instantaneous configurations that can be quantified by measuring how the pinning of randomly selected atoms on the average restraint. Analogous studies of the impact of pinning on relaxation kinetics have been previously reported [20]. Let $M = (1-c)N$ be the number of unpinned particles where c is the number fraction of pinned particles. Following ref. [15], the impact of pinning on the instantaneous normal modes is calculated by diagonalizing the $3M \times 3M$ Hessian matrix consisting of only derivatives of coordinates of the unpinned particles. At no point do we impose pinning constraints in the simulations used to generate the configurations, only in our normal mode analysis of those configurations.

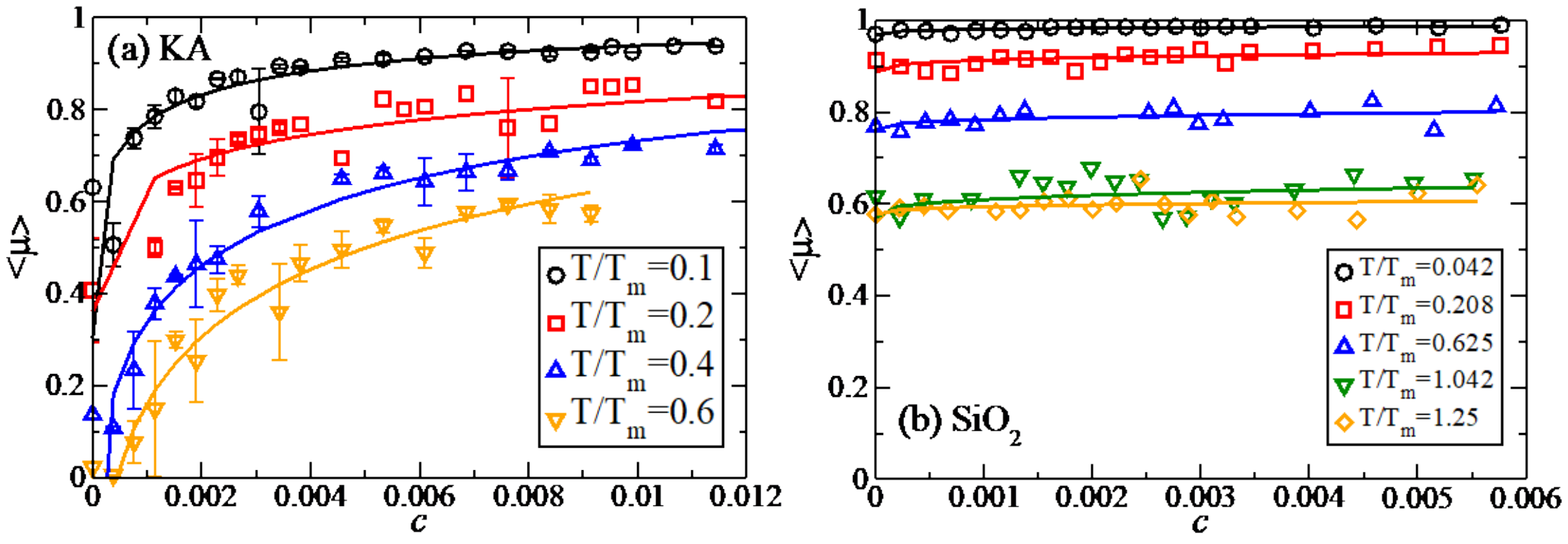

**Figure 4.** The average restraint <μ> as a function of the pinning concentration c for a) KA mixture and b) $SiO_2$ for a range of temperatures as shown. The solid curves are fits to Eq. 3.

In Fig. 4 we plot the dependence of <μ> as a function of the pinning concentration c for a range of temperatures. We find that the fragile liquid exhibits a substantial increase in restraint as a result of the pinning of a small fraction of particles while silica is largely insensitive to the pinning, being inherently highly restrained. As pinning increases the restraint of the fragile liquid, it inevitably shifts the structure of the liquid, as measure by restraint, towards that of the strong liquid and, hence, decreases its fragility.

Previously [15], we found that variation in <μ> could be described by

$$<\mu(c)>=1-(1-<\mu(0)>)\exp\left(-\xi c^{1/3}\right) \qquad (3)$$

where $\xi^3$ is roughly the number of unpinned particles restrained by each pinned particle. ($\xi$ can be regarded as an effective length in units of a characteristic particle dimension.). The curves in Fig. 4 are the fits to Eq. 3 with the fitted values of $\xi^3$ plotted in Fig. 5 as a function of $T/T_m$ for the two liquids. We find the impact of pinning to be considerably larger in the fragile liquid than the strong, a direct consequence of the lower restraint in the unpinned fragile liquid. This pinning extent is entirely mechanical in origin since, by construction, the different pinning concentrations only change the normal modes analysis, not the configuration itself. The increase in $\xi^3$ on cooling reflects contributions from both the increase in restraint of the unpinned liquid and the generic consequence of pinning on a harmonic system of particles. The increase in the silica $\xi^3$ at low temperatures appears to be an example of the latter process. Resolving these contributing effects we leave to future work.

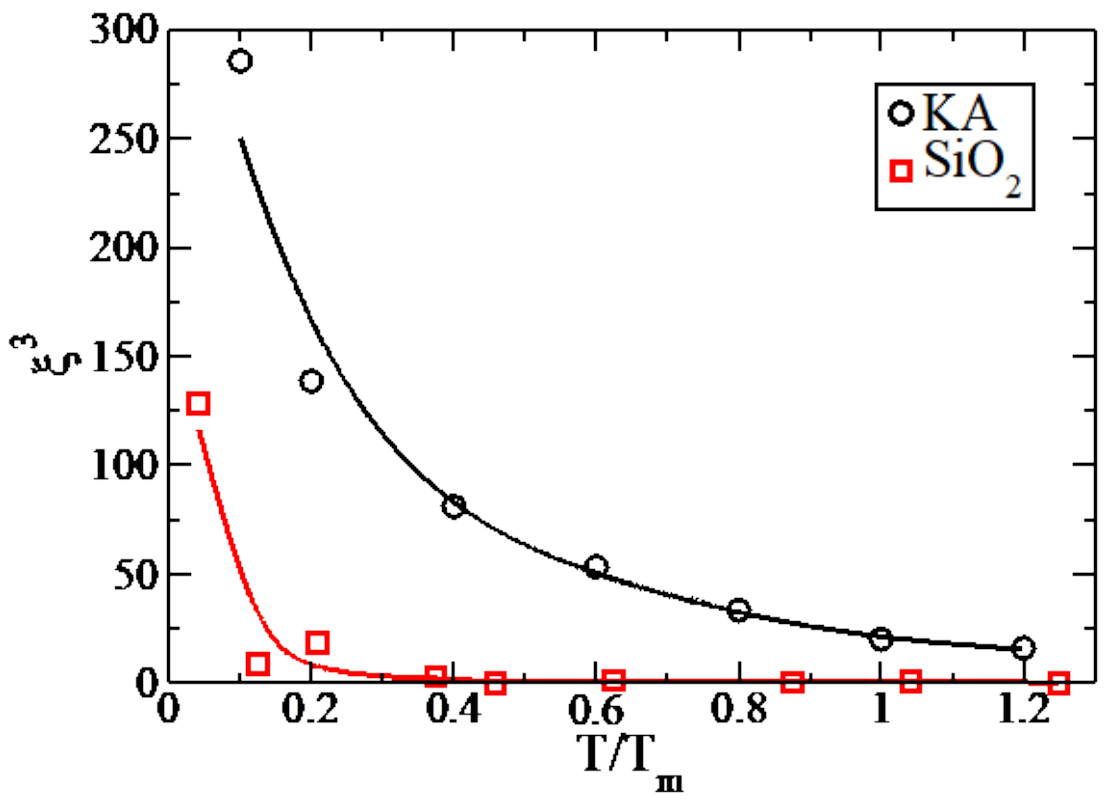

**Figure 5.** The number unpinned particles restrained by each pinned particle, $\xi^3$, is plotted as a function of $T/T_m$ for the KA mixture and $SiO_2$. These values are extracted from the fits to Eq.3. The curves are spline fits to guide the eye.

## 4. Fragility and the Potential Energy Landscape (PEL)

There is a considerable and diverse literature regarding features of the PEL that determine the fragility of the liquid. In 1995, Angell [3] argued that fragility increased with the density of local minima per unit energy. This proposal was qualified by Ruocco *et al* [21] who found that the related configurational entropy was insufficient to determine the fragility in model calculations. Stillinger [22] suggested that the PEL of a fragile liquid contained multiple metabasins, imagined as large amplitude extended undulations in the potential energy surface on which smaller scale local minima were superimposed. A strong liquid, in contrast, existed within a single metabasin. Heuer and coworkers [23,24], in exploring a computational realization of the metabasin concept, concluded that fragility was determined by the shape of the local minima energy distribution at low energies. In silica [24], they noted, this density of states ends discontinuously, so that there is a considerable number of minima with the lowest energy (i.e. the CRN structure [12]). In contrast, the number of minima in the fragile liquid decreases continuously to zero as the energy decreases. Coslovich and Pastore [25] found a correlation between the temperature dependence of the average barrier height, estimated using a method proposed by Cavagna [26], and the fragility of different Lennard-Jones mixtures.

A direct benefit of establishing a structural difference between a strong and a fragile liquid is that the connection between structure and energetics is considerably more straightforward to establish than that between the relaxation kinetics and energy. We can begin by replotting, in Fig. 6, the values of <μ> against the potential energy of the configuration. In order to compare our two liquids, we shall consider a reduced energy, $\frac{U-U_{min}}{U_{max}-U_{min}}$ where $U_{min}$ and $U_{max}$ are the minimum and maximum values of the potential energy observed over all temperatures. The measure essentially represents the fractional 'height' of the configuration energy in the PEL. As the restraint μ is a monotonically increasing function of the average local curvature of the potential, the curves in Fig. 6 can be read as measure of the average local curvature, relative to $k_BT$, of the PEL as a function of how far down into the energy landscape we are. Viewed this way, the structural difference between the PEL's of the strong and fragile liquids arises, not from the distribution of local minima energies, but from the degree to which the liquid accesses low (or negative) curvatures as its approaches the top of the landscape. In a strong liquid, the fraction of configurations sampling negative curvatures at high energies is very much smaller than that of a fragile liquid. This picture makes no mention of the distribution of barrier heights that are conventionally invoked to account for the kinetic manifestations of fragility

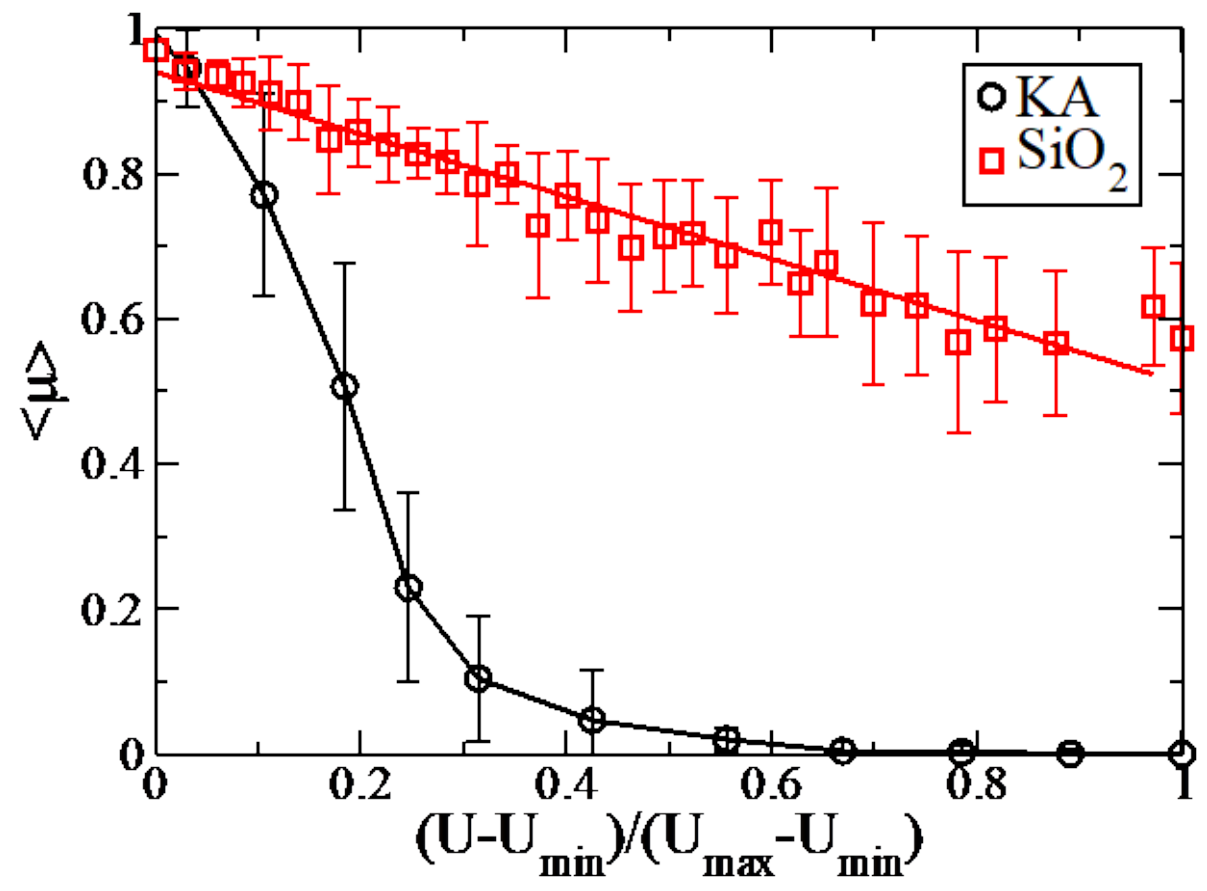

**Figure 6.** A plot of <μ> for the KA mixture and silica as a function of a reduced potential energy $(U-U_{min})/(U_{max}-U_{min})$ where $U_{max}$ = <U> at the maximum T studied (i.e. the top of the landscape) and $U_{min}$ = <U> at T=0.

**5. Conclusion**

In this paper we demonstrate that a simple and quite general measure of the degree of restraint experienced by each atom in an instantaneous configuration differentiates a strong liquid from a fragile one without requiring any reference to the temperature dependence of relaxation kinetics. Structurally, the strong liquid differs from the fragile in that its configurations exhibit a high degree of restraint over its entire temperature range. The fragile configurations, in contrast, become unrestrained at high temperature. The difference between strong and fragile liquids is, according to our result, most striking at high temperature. It is here that the fragile liquid's restraint is zero and its susceptibility to pinning is high while the strong liquid is the inverse: high restraint and low pinning susceptibility.

This structural account may prove useful in explaining the observed difference in relaxation kinetics (e.g. a 'stiffer' liquid will have larger activation energies, higher probability of reversals, etc). The immediate benefit of this structural picture, however, is that we can address interesting questions about supercooled liquids without having to refer to dynamics at all. What, exactly, are these interesting questions? In this paper we presented measures of the difference in structural heterogeneity of the two liquids and quantified the difference in the susceptibility of restraint to the pinning of random particles. Here we offer up a number of suggestions for future study.

1) How is the structural fragility of network melts like silica adjusted down by the addition of low valence ions? Here we refer to the details of the microscopic relaxation of restraint achieved by such additives.

2) How is the structural fragility influenced by the proximity to a free surface – both at equilibrium and during deposition from the gas phase or from solution?

3) In considering the rheology of liquids with different fragilities, the order parameter described here can be used to explore how the microscopic distribution of streaming velocities correlate with the spatial variations in microscopic restraint. Here we would make use of the fact that the restraint order parameter is defined on instantaneous configurations and so can be use in shearing fluids as easily as equilibrium ones.

The fundamental question regarding fragility is to identify the general features of the Hamiltonian that differentiated the strong from the fragile liquid. In this paper, we find that the essential difference in the energy landscapes that determine fragility is not the distribution of well depths or barrier heights often cited. Instead, our result is that it is the degree to which configurations can access regions of low or negative curvature of the underlying PEL at high energies. It is the fragile liquids access to these low curvature regions that leads to the inability of high energy configurations to impose restraint on the motions of the constituent atoms. Conversely, liquid silica rarely strays from regions of high positive curvature, even when approaching the boiling point.

In a 2008 paper [27], Austen Angell laid out a “Big Picture” account of glass forming liquids, one that assigned fragility to the existence of a sub-$T_m$ cooperative (discontinuous) thermodynamic transition and the loss of fragility with the weakening of the transition and its shift to higher temperatures. In this paper we have presented a structural account of fragility that aligns with at least a part of this picture. While the change in atomic restraint arises as a direct mechanical consequence of the PEL rather than through a thermodynamic transition [15], we do find that different fragilities are indeed linked through a single coherent picture, one that includes sub-$T_m$ structural transformation for fragile liquids that, in silica, has been shifted, at least nominally, to a temperature above the boiling point. The further testing of this picture through the study of atomic restraint in alkali substituted silicates of varying fragility would provide insight into how the underlying structural transformation evolves between the two limiting cases presented here.

**Acknowledgements**

PH gratefully acknowledges an unfailingly enjoyable exchange of ideas with Austen Angell over many years. The authors acknowledge support from the Australian Research Council.

**References**

1. W.T.Laughlin and D.R.Uhlmann, Viscous flow in simple organic liquids. J. Phys. Chem. **76**, 2317-2325 (1972).

2. C. A. Angell, Spectroscopy simulation and scattering, and the medium range order problem in glass. J. Non-Cryst. Sol. **73**, 1-17 (1985).

3. C. A. Angell, Formation of glasses from liquids and biopolymers. Science **267**, 1924-1935 (1995).

4. P. G. Debenedetti and F. H. Stillinger, Supercooled liquids and the glass transition. Nature **410**, 259-267 (2001).

5. A. Heuer, Exploring the potential energy landscape of glass-forming systems: from inherent structures via metabasins to macroscopic transport. J. Phys.: Cond. Matt. **20**, 373101 (2008).

6. A. Widmer-Cooper, H. Perry, P. Harrowell and D. R. Reichman, Irreversible reorganization in a supercooled liquid originates from localized soft modes. *Nature Phys*. **4**, 711-715 (2008).

7 J. Ding, S. Patinet, M. Falk, Y. Cheng and E. Ma, Soft spots and their structural signature in a metallic glass. *Proc. Nat. Acad. Sci USA* **111**, 14052-14056 (2014).

8. W. Kob and H. C. Andersen, Testing mode-coupling theory for a supercooled binary Lennard-Jones mixture – The van Hove correlation function. Phys. Rev. **51**, 4626-4641 (1995).

9. D. Coslovich and G. Pastore, Understanding fragility in supercooled Lennard-Jones mixtures: I. Locally preferred structures. J. Chem. Phys. **127**, 124504 (2007).

10. C. P. Royal and S. R. Williams, The role of local structure in dynamical arrest, Phys. Rep. **560**, 1-75 (2015).

11. D. Wei, J. Yang, M.-Q. Jiang, L.-H. Dai, Y.-J. Wang, Dyre, I. Douglass and P. Harrowell, Assessing the utility of structure in amorphous materials. J. Chem. Phys. **150**, 114502 (2019).

12. W. H. Zachariasen, The atomic arrangement in glass. J. Am. Chem. Soc. **54**, 3841-3851 (1932).

13. R. G. Della Valle and H.C. Andersen, Molecular dynamics simulation of silica liquid and glass. J. Chem. Phys. **97**, 2682 (1992).

14. K. Vollmayr, W. Kob and K. Binder, Cooling-rate effects in amorphous silica: a computer-simulation study. Phys. Rev. B **54**, 15808-15827 (1996).

15. G. Sun and P. Harrowell, A General Structural Order Parameter for the Amorphous Solidification of a Supercooled Liquid, preprint http://arxiv.org/abs/2109.12461 (2021).

16. T. S. Grigera and G. Parisi, Fast Monte Carlo algorithm for supercooled soft spheres. Phys. Rev. E **63**, 045102 (2001).

17. S. Sundararaman, L. Huang, S. Ispas, W. Kob, New optimization scheme to obtain interaction potentials for oxide glasses. J. Chem. Phys. **148**, 194504 (2018).

18. U. R. Pedersen, T. B. Schroder and J. C. Dyre, Phase diagram of Kob-Andersen-type binary Lennard-Jones mixtures. Phys. Rev. Lett. **120**, 165501 (2018).

19. A. Takada, P. Richet, C. R. A. Catlow and G. D. Price, Molecular dynamics simulations of vitreous silica structures. J. Non-Cryst. Solids **345-346**, 224-229 (2004).

20. S. Chakrabarty, S Karmakar and C. Dasgupta, Dynamic of glass forming liquids with randomly pinned particles. Sci . Rep. **5**, 12577 (2015).

21. G. Ruocco, F. Sciortino, F. Zamponi, C. De Michele and T. Scopigno, Landscapes and fragilities. J. Chem. Phys. **120**, 10666-10680 (2004).

22. F. H. Stillinger, A topographic view of supercooled liquids and glass formation. Science **267**, 1935-1939 (1995).

23. B. Doliwa and A. Heuer, Energy barriers and activated dynamics in a supercooled Lennard-Jones liquid. Phys. Rev. E **67**, 031506 (2003).

24. A . Saksaengwijit and A. Heuer, Dynamics of liquid silica as explained by properties of the potential energy landscape. Phys. Rev. E **73**, 061503 (2006).

25. D. Coslovich and G. Pastore, Understanding fragility in supercooled Lennard-Jones mixtures: II. Potential energy surface. J. Chem. Phys. **127**, 124505 (2007).

26. A. Cavagna, Fragile vs strong liquid: a saddle-rules scenario. Europhys. Lett. **53**, 490-496 (2001).

27. C. A. Angell, Glass-formers and viscous liquids since David Turnbull: enduring puzzles and new twists. MRS Bull. **33**, 544-555 (2008)